\documentclass[prl,aps,twocolumn,floats,showpacs]{revtex4}
\usepackage{graphicx}
\usepackage{psfrag}

\begin{document}

\title{Do all pure entangled states violate Bell's inequalities for correlation functions?}

\author{Marek {\. Z}ukowski$^1$, {\v C}aslav Brukner$^2$, Wies{\l}aw Laskowski$^3$
and Marcin Wie{\'s}niak$^3$}
\affiliation{$^1$Instytut Fizyki Teoretycznej i Astrofizyki, Uniwersytet
Gda\'nski, PL-80-952 Gda\'nsk, Poland\\ $^2$Institut f\"ur
Experimentalphysik, Universit\"at Wien,  Boltzmanngasse 5, A--1090
Wien, Austria\\ $^3$Wydzia{\l } Matematyki i Fizyki, Uniwersytet
Gda\'nski, PL-80-952 Gda\'nsk, Poland}

%\date{\today}

\begin{abstract}

Any pure entangled state of two particles violates a Bell
inequality for two-particle correlation functions (Gisin's
theorem). We show that there exist pure entangled $N\!>\!2$ qubit
states that do not violate any Bell inequality for $N$ particle
correlation functions for experiments involving two dichotomic
observables per local measuring station. We also find that
Mermin-Ardehali-Belinskii-Klyshko inequalities may not always be
optimal for refutation of local realistic description.

\end{abstract}


\maketitle

Quantum mechanics violates Bell type inequalities
\cite{bell,chsh,ghz,mermin}, which hold for any local realistic
theory. In a realistic model the measurement results are
determined by "hidden" properties the particles carry prior to and
independent of observation. In a local model the results obtained
at one location are independent of any measurements or actions
performed at space-like separation. The theorem of Gisin
\cite{gisin} states that {\it any} pure non-product state of two
particles violates a Clauser-Horne-Shimony-Holt (CHSH) \cite{chsh}
inequality, which involves only two-particle correlation
functions, for two alternative dichotomic measurements for each of
the local observers.

Can Gisin's theorem be generalized to all $N$-particle pure
entangled states? We show here that this is not the case for Bell
inequalities involving only correlation functions in experiments
in which local observers can choose between two dichotomic
observables. We find a family of pure entangled states of $N$
qubits which do not violate {\it any} such Bell inequality. This
family is a subset of a larger one of the generalized GHZ states
given by
\begin{equation}
|\psi\rangle =
\cos\alpha|0,...,0\rangle+\sin\alpha|1,...,1\rangle. \label{STATE}
\end{equation}
with $0\!\leq \!\alpha \!\leq\! \pi/4$. The GHZ states \cite{ghz} are for $\alpha=\pi/4$

Scarani and Gisin \cite{scarani} noticed a surprising feature of
such states. They show that for $\sin{2\alpha}\! \leq \!
1/\sqrt{2^{N-1}}$ the states (\ref{STATE}) {\it do not violate}
the Mermin-Ardehali-Belinskii-Klyshko (MABK) inequalities
\cite{mermin}. This has been obtained numerically for
$N\!=\!3,4,5$ and conjectured for $N\!>\!5$. Their result
contrasts the case of $N\!=\!2$ of two qubits and is surprising as
the states (\ref{STATE}) are a generalization of the GHZ states
\cite{ghz} which violate maximally the MABK inequalities.

Scarani and Gisin write that ''this analysis suggest that MK
[here, MABK] inequalities, and more generally the family of Bell's
inequalities with two observables per qubit, may not be the
'natural' generalizations of the CHSH inequality to more than two
qubits'' \cite{scarani}.  Concerning this question we find here an
interesting discrepancy between the case of even and odd number of
qubits. We prove that for all $N$ {\it odd} and for
$\sin{2\alpha}\leq 1/\sqrt{2^{N-1}}$ the generalized GHZ
states satisfy {\it all} possible Bell inequalities for
$N$-particle correlation functions, which involve two alternative
dichotomic observables at each local measurement station. Note
also, that since the reduced density matrices of all proper
subsystems of the $N$-qubit system described by (\ref{STATE}) are
separable, no Bell inequality for $K\!<\!N$ particle correlation
functions can be violated. Thus, Gisin's theorem cannot be
straightforwardly generalized in this case. We also find that for
all $N$ {\it even} the generalized GHZ state {\it always} violates
a Bell inequality for $N$-particle correlation functions.
Interestingly, this inequality, as conjectured in \cite{scarani},
is not a MABK one.  This implies that MABK inequalities may not
always be optimal for refutation of local realism.

It is important to notice that the measurements on qubits in the
states (\ref{STATE}) can violate Bell inequalities
%of a different kind,
if (i) one makes an additional postselection of measurements,
and/or (ii) not all $N$ observers are separated. Popescu and
Rohrlich \cite{popescu} showed that no local realist description
is possible for any pure multipartite entangled state, provided
additional manipulations are allowed. Specifically, for the
state (\ref{STATE}) a pair of observers can perform an analysis of
their correlations conditioned on a specific set of results of
other $N\!-\!2$ observers. This can lead to a violation of a CHSH
inequality for the pair of observers. For example, each of
$N\!-\!2$ observers measures his qubits in the basis $|0'\rangle=
1/\sqrt{2}(|0\rangle+ |1\rangle)$ and
$|1'\rangle=1/\sqrt{2}(|0\rangle - |1\rangle)$. Then if all of
them obtain result $0'$, the first two observers are left with the
state $|\psi\rangle =
\cos\alpha|0,0\rangle+\sin\alpha|1,1\rangle$, which always
violates a CHSH inequality \cite{gisin}. If (ii) is allowed, then
the $N$ observers can be separated into two spatially separate
groups of $M$ and $N\!-\!M$ observers, which make joint
measurements. For both groups one can always find dichotomic
observables, such that a CHSH inequality for correlations between
the measurements of the two groups is violated. Simply, the state
(\ref{STATE}) can be considered as a ''two-qubit'' state
$|\psi\rangle = \cos\alpha|0_M 0_{N-M}\rangle+\sin\alpha|1_M
1_{N-M}\rangle$ where $|0_M\rangle$, $|1_M\rangle$, and
$|0_{N-M}\rangle$, $|1_{N-M}\rangle$ are states of the qubits of
the group with $M$ and with $N-M$ observers, respectively. Thus
Gisin's theorem applies. Note also that, Bennett {\it et al.}
\cite{bennett} showed that the states (\ref{STATE}) are reversibly
generated from the GHZ states and D\"{u}r \cite{duer} proved that all
pure entangled states of $N$ qubits are distillable. However, in
the present paper, as in the standard Bell-GHZ type situation, we
are interested in the correlations between measurements on $N$
{\em separated qubits without postselection}.

We base our analysis on the results of Ref.\cite{zukowski}. Here
is a summary. A single generalized Bell inequality was derived,
which is equivalent to the full set of $2^{2^N}$ Bell inequalities
for the correlation functions between measurements on $N$
particles, which involve two alternative dichotomic observables at
each local measurement station \cite{werner,zukowski,weinfurter}.
The correlation functions for such measurements cannot be
described by a local realistic model, if and only if, the
generalized Bell inequality is violated. In parallel, in Ref.
\cite{zukowski} a {\it necessary} and {\it sufficient} condition
was derived for the correlations of $N$ qubits in an arbitrary
state to violate the generalized Bell inequality. (Recently it was
shown \cite{acin}, that all states violating the generalized Bell
inequality are distillable.)

The condition is as follows. Consider N observers and allow each
of them to choose a local coordinate system. An arbitrary quantum
state $\rho$ of $N$ qubits can be represented by tensor products of local
Pauli operators in the following way
\begin{equation}
\rho=\frac{1}{2^N} \sum_{x_1,...,x_N=0}^{3} T_{x_1...x_N} \mbox{ }
\sigma_{x_1} \otimes ... \otimes \sigma_{x_N}. \label{state}
\end{equation}
Here $\sigma_0$ is the identity operator and $\sigma_{1}$,
$\sigma_{2}$ and $\sigma_{3}$ are three Pauli operators associated
with mutually orthogonal directions. The coefficients
$T_{x_1...x_N} =\mbox{Tr}[\rho(\sigma_{x_1} \otimes ... \otimes
\sigma_{x_N})]$ for $x_i=1,2,3$ are elements of the $N$-qubit
correlation tensor $\hat{T}$.

The {\it necessary and sufficient condition for a quantum state to
satisfy the generalized $N$-particle Bell inequality} is given by
the following criterion: in {\it any} set of local coordinate
systems for the $N$ observers and for any set of unit vectors
$\vec{c}^j=(c^j_1,c^j_2)$ one has
\begin{equation}
T^{mod}_{c^1...c^N} \equiv \sum_{x_1,...,x_N= 1}^{2} c^1_{x_1} ...
c^N_{x_N} \mbox{ } |T_{x_1...x_N}| \leq 1. \label{mod}
\end{equation}
This can be restated as follows. Suppose, one replaces the
components of the correlation tensor $T_{x_1...x_N}$ by their
moduli $|T_{x_1...x_N}|$, and builds of such moduli a new tensor
$\hat{T}^{mod}$. Suppose moreover, that this new tensor is
transformed to a new set of local coordinate systems.  This
transformation is for each local observer a rotation in the plane
spanned by the orthogonal directions 1 and 2. If the coordinates
of the transformed tensor $T^{mod}_{c^1...c^N}$ always satisfy
constraint (\ref{mod}), i.e. are not greater than $1$, then, and
only then, a local realistic description is possible.

By applying the Cauchy inequality to the middle term of expression
(\ref{mod}) one obtains a simple {\it sufficient condition for the
generalized $N$-particle inequality to be satisfied}: in {\it any}
set of local coordinate systems of $N$ observers one must have
\begin{equation}
\sum_{x_1,...,x_N = 1}^{2} T^2_{x_1...x_N} \leq 1. \label{ban}
\end{equation}

In the conditions (\ref{mod}) and (\ref{ban}) the sums are taken
over any two orthogonal axes 1 and 2 for all local coordinate
systems. The two axes can be either $x$ and $y$, or $x$ and $z$,
or $x$ and $z$. Since the full group of rotations contains within
itself cyclic permutations of the coordinates (e.g $x\! \to\! y
\!\to \!z\! \to\! x$), it is enough to consider just one
particular choice for a set pairs of local axes, and then consider
{\it arbitrary} local transformations of the full tensor. In other
words, one can work with  a particular sub-tensor of the
correlation tensor, which is build by the components containing
only, say, the $x$ and $y$ indices.

Once a sub-tensor is chosen, then on one hand, if the inequality
(\ref{ban}) holds under {\it arbitrary} changes of the local
coordinate systems, the correlations described by the quantum
state satisfy the generalized inequality. On the other hand, if
the condition (\ref{mod}) is violated at least for one choice of
local coordinate systems, no local realistic description is
possible for the $N$-particle correlations.

When considering arbitrary rotation, we will use the Euler
theorem. Any rotation of a local coordinate system can be
expressed as a sequence of rotations around three axes $z$, $x'$
and $z''$. Note that rotations of coordinate systems of different
observers commute.

Using a lengthy but otherwise straightforward algebra one can show
that the correlation tensor for the states (\ref{STATE}) has only
the following nonvanishing components: (i) For $N$ even $T_{z...z}=1$ and
$T_{x...x}=\sin2\alpha$, and all components with $2k$ $y$'s and
otherwise only $x$'s (e.g., for $N=4$, $T_{xyxy}$, etc.) are equal
to $(-1)^k \sin2\alpha$. (ii) For $N$ odd $T_{z...z}=\cos2\alpha$, and
$T_{x...x}=\sin2\alpha$, and all components with $2k$ $y$'s and
otherwise only $x$'s (e.g., for $N=3$, $T_{xyy}$, etc.) are equal
to $(-1)^k \sin2\alpha$. It will be convenient to represent the correlation
tensor as a sum of tensor products of unit three-dimensional vectors:
\begin{equation}
\hat{T} = \sum_{x_1,...,x_N=1}^{3} T_{x_1...x_N} \vec x_1\!
\otimes\! ... \!\otimes\! \vec x_N.
\end{equation}

We now give the main technical results.

Statement 1: {\it For $\sin2\alpha\leq 1/\sqrt{2^{N-1}}$ and $N$
odd, the correlations between measurements on qubits in the
generalized GHZ state (\ref{STATE}) satisfy all Bell inequalities
for correlation functions, which involve two dichotomic
observables per local measurement station.}

The main idea of the proof is to show that condition (\ref{ban})
is satisfied for the range of $\alpha$ given above. We give the
proof for $N=3$. We show that, for this range of $\alpha$, the
value of $\sum_{i,j,k=x,y} T_{ijk}^2$ after an arbitrary set of
local rotations is performed is never larger than one. The proof
for general odd $N$ is a straightforward generalization.

The correlation tensor of state (\ref{STATE}), for $N=3$, reads
\begin{eqnarray}
\hat T \! &=& \! \cos{2 \alpha} ~ \vec z_1 \! \otimes\! \vec z_2
\!\otimes\! \vec z_3 + \sin{2 \alpha} \Big[ \vec x_1 \!\otimes
\!\vec x_2\! \otimes\! \vec x_3 \label{t} \\ &-& \vec x_1
\!\otimes \!\vec y_2 \!\otimes\! \vec y_3 - \vec y_1 \!\otimes\!
\vec x_2\! \otimes \!\vec y_3 - \vec y_1 \!\otimes\! \vec y_2
\!\otimes\! \vec x_3 \Big]. \nonumber
\end{eqnarray}

We first rotate the local coordinate systems of each of the
observers around the local $\vec z$ axes. Such a set of rotations
of course leaves  $\sum_{i,j,k=x,y}  T_{ijk}^2$ invariant,
therefore it stays put at its initial value $4\sin^2{2\alpha}\leq
1$. The correlation tensor in the new set of local coordinate systems
is given by
\begin{equation}
\hat{T'} = \cos{2 \alpha} ~ \vec z_1 \otimes \vec z_2
\otimes \vec z_3 \label{st} \\ + \sin{2 \alpha} \hspace{-0.2cm}
\sum_{i,j,k=x,y} T'_{ijk} \vec i'_1 \otimes\vec j'_2
\otimes \vec k'_3 \nonumber
\end{equation}
The  values for $T'_{ijk}$ can be obtained by replacing $\vec
x_i\to\cos{\phi_i}\vec x'_i+\sin{\phi_i}\vec y'_i$ and $\vec
y_i\rightarrow\cos{\phi_i}\vec y'_i-\sin{\phi_i}\vec x'_i$ in Eq.
(\ref{t}) for $\hat T$. The components satisfy the following
relations
\begin{eqnarray}
T'^2_{xxx}\!&=&\!T'^2_{yyx}\!=\!T'^2_{yxy}\!=\!T'^2_{xyy}\!=\!T^2_{xxx}
\cos^2{\sum_{i=1}^3\phi_i}, \\
T'^2_{yxx}\!&=&\!T'^2_{xyx}\!=\!T'^2_{xxy}\!=\!T'^2_{yyy}\!=\!T^2_{xxx}
\sin^2{\sum_{i=1}^3\phi_i}.
\end{eqnarray}

Next, we rotate the local coordinate systems of each of the
observers around the local $\vec x'$ by the angle $\theta_i$. Now
the specific values for $T^{''}_{ijk}$ can be obtained by
replacing $\vec z'_i \rightarrow \cos{\theta_i} \vec z''_i +
\sin{\theta_i} \vec y''_i $ and $\vec y'_i \rightarrow
\cos{\theta_i} \vec y''_i - \sin{\theta_i} \vec z''_i$ in Eq.
(\ref{st}). The new components in the $xy$ sector of the
correlation tensor satisfy the following relations
\begin{eqnarray}
T''^2_{xxx} \! &=& \! T'^2_{xxx}, \hspace{0.8cm}  T''^2_{xxy}
\!=\! c^2_3 T'^2_{xxy}, \hspace{0.6cm} T''^2_{xyx} \!=\! c^2_2
T'^2_{xyx} \label{sapat} \\ T''^2_{yxx} \!&=&\! c^2_1 T'^2_{yxx},
\hspace{0.5cm} T''^2_{xyy}\! =\! c^2_2 c^2_3 T'^2_{xyy},
\hspace{0.3cm} T''^2_{yxy}\! =\! c^2_1 c^2_3 T'^2_{yxy}
\nonumber \\ \nonumber T''^2_{yyx} \!&=& \!c^2_1 c^2_2 T'^2_{yyx}, \hspace{0.2cm}
T''^2_{yyy} \!=\! (T_{zzz} s_1 s_2 s_3 + T'_{yyy} c_1 c_2
c_3)^2,\\ \nonumber
\end{eqnarray}
where $s_i \equiv \sin{\theta_i}$ and $c_i \equiv \cos{\theta_i}$.
By applying the Cauchy inequality $ (Ac_1+Bs_1)^2\leq A^2+B^2$ to
the last relation in (\ref{sapat}), we obtain that $T''^2_{yyy}
\leq T_{zzz}^2 s_2^2 s_3^2 + T^{'2}_{yyy} c_2^2 c_3^2$.

Therefore we obtain for the sum
\begin{eqnarray}\hspace{-0.3cm}  
\sum_{i,j,k=x,y} \hspace{-0.3cm}  T^{''2}_{ijk} & \!\leq \! &  c_1^2
c_2^2 T^{'2}_{yyx} + c_2^2 c_3^2 T^{'2}_{xyy} + c_1^2 c_3^2
T^{'2}_{yxy}+ c_1^2 T^{'2}_{yxx} \nonumber \\ \hspace{-0.2cm} &\!+\!&
c_2^2 T^{'2}_{xyx} + c_3^2 T^{'2}_{xxy} + T^{'2}_{xxx} + T_{zzz}^2
s_2^2 s_3^2 + T^{'2}_{yyy} \nonumber c_2^2 c_3^2.
\end{eqnarray}
This expression is a linear function of its arguments $c_1^2,
c_2^2,$ and $c_3^2$. Thus, its maximal value is reached at the
border of the region for which the function is defined, i.e. at
$c_i^2 =$ 0 or 1. For $c_1\!=\!c_2\!=\!c_3\!=\!1$ one has
\begin{equation}
\sum_{i,j,k=x,y} T^{''2}_{ijk} \leq 4\sin^2 2\alpha
\end{equation}
and for all other cases $\sum_{i,j,k=x,y}T^{''2}_{ijk} \leq 1$
provided $\sin2\alpha\leq 1/2$. Thus, for the considered range of
$\alpha$, after arbitrary subsequent local rotations along three local $z$
axes and arbitrary local rotations along three $x'$ axes, one has
$\sum_{i,j,k=x,y} T^{''2}_{ijk}\leq 1$.

The final stage of our proof rest upon the observation, that the
final Euler rotation of the local coordinate systems of each of
the three observers around axes $ z''$ leaves the sum of squares
of the components in the $xy$ sector of the tensor {\it
invariant}.

The proof for an arbitrary odd $N$ follows  the same pattern. For
$\alpha$ satisfying $\sin{2\alpha}\leq 1/\sqrt{2^{N-1}}$ the
generalized Bell inequalities cannot be violated. The threshold
value for $\alpha$ decreases exponentially with the growing odd
$N$ because the number of nonvanishing components in the $xy$
sector of the correlation tensor of the state (\ref{STATE}) is
$2^{N-1}$.

Thus the sufficient condition is met. All possible Bell
inequalities for $N$ particle correlation functions  in tests
involving two alternative dichotomic observables for each observer
must hold, for the range of $\alpha$ given above. Recalling that
for (\ref{STATE}) any subset of less than $N$ qubits is in a
separable state, we conclude that {\it all} possible Bell
inequalities for correlation functions in tests involving two
alternative observables for each observer must hold, for the range
of $\alpha$ given above. Outside of this range the MABK
inequalities can always be violated \cite{scarani}, and
consequently also the generalized Bell inequality.

{Statement 2:} {\it The correlations between measurements for an
{\it even} number of qubits in the generalized GHZ state
(\ref{STATE}) cannot be described within a local realistic model.}

We shall show, that for all even $N$ the generalized GHZ state
violates the constraint (\ref{mod}). We give the proof only for
$N=4$. The generalization to arbitrary $N$ is obvious.
Of course the $N=2$ is covered by Gisin's theorem \cite{gisin}.

For $N=4$ the sector of the correlation tensor $\hat T$, which is
limited to $x$ and $z$ components, is given by
\begin{equation}
\hspace{-0.05cm} \hat T_{[zx]} \!\equiv\! \vec z_1 \otimes \vec
z_2 \otimes \vec z_3 \otimes \vec z_4 + \sin{2\alpha}
\hspace{0.1cm} \vec x_1 \otimes \vec x_2 \otimes \vec x_3 \otimes
\vec x_4. \label{cep}
\end{equation}
Let us first rotate the coordinate axes of the first three
observers around the local directions $\vec y$ by $45^o$, that is
$\vec z_i = \frac{1}{\sqrt{2}} (\vec z'_i + \vec x'_i)$ and $\vec
x_i = \frac{1}{\sqrt{2}} (\vec x'_i - \vec z'_i)$ for $i=1,2,3$.
The sub-tensor in the new coordinates reads
\begin{eqnarray} \hat T'_{[zx]} &=&   2^{-3/2}
\Big[ (\vec x'_1 + \vec z'_1) \otimes (\vec x'_2 + \vec z'_2)
\otimes (\vec x'_3 + \vec z'_3) \\ &\otimes& \vec z_4'  +
\sin{2\alpha} (\vec x'_1 - \vec z'_1) \otimes (\vec x'_2 - \vec
z'_2) \otimes (\vec x'_3 - \vec z'_3) \otimes \vec x_4' \Big]
\nonumber
\end{eqnarray}
Next, we build a new tensor $\hat T^{'(mod)}_{[zx]}$ by replacing
all components of $\hat T'_{[xz]}$ by their moduli:
\begin{eqnarray}
\hat T^{'(mod)}_{[zx]} &=& 2^{-3/2} (\vec x'_1 + \vec z'_1)
\otimes (\vec x'_2 + \vec z'_2) \nonumber \\ &\otimes& (\vec x'_3
+ \vec z'_3) \otimes (\vec z'_4 + \sin{2\alpha} \vec x'_4).
\end{eqnarray}
Finally, we change the local coordinate systems of the observers
in the following way: $\vec x''_i = \frac{1}{\sqrt{2}}(\vec x'_i +
\vec z'_i)$ for $i=1,2,3$ and $\vec x''_4 = \frac{1}{\sqrt{1+
\sin^2{2\alpha }}} (\vec z_4' + \sin{2\alpha} \vec x_4')$. The
$xz$ sector of the tensor in the new coordinates reads
\begin{equation}
\hat T^{''(mod)}_{[zx]} = \sqrt{1+ \sin^2{2\alpha}} ~ \vec x''_1
\otimes \vec x''_2 \otimes \vec x''_3 \otimes \vec x''_4
\end{equation}
Therefore we find local coordinate systems in which our modified
tensor has a component of a value higher than 1 for all
non-vanishing values of $\sin{2\alpha}$. The criterion (\ref{mod})
is violated. No local realistic description is possible.

To generalize the proof to an arbitrary even $N$ it is enough  to
notice, that the sub-tensor built out of the $zx$ components of
the full tensor has the characteristic form $$\hat T_{[zx]} = \vec
z_1 \otimes \vec z_2 ... \otimes \vec z_N + \sin{2\alpha}
\hspace{0.1cm} \vec x_1 \otimes \vec x_2 ....\otimes \vec x_N,$$
and proceed as in the case $N=4$.

The correlations between the measurements on an even number of
qubits in the GHZ generalized state do not allow any local
realistic model. Since the generalized Bell inequality is
violated, i.e. at least one out of the full set of $2^{2^N}$
\cite{werner,zukowski} inequalities implied by the generalized one
must be violated. Surprisingly, the violated inequality is not a
MABK one but a generalized CHSH inequality.

Consider a Bell experiment in which observers 1 and 2 choose
between two dichotomic observables and the other $N\!-\!2$ ones
keep their settings at $\vec{z}$ unchanged. The states $|0\rangle$
and $|1\rangle$ in Eq. (\ref{STATE}) are eigenstates of
$\vec{z}\cdot\vec{\sigma}$. Since $N\!-\!2$ is also even, the product
of the local results of the $N\!-\!2$ observers for the case of
the generalized GHZ state (\ref{STATE}) is always $1$. Thus their
results effectively do not contribute to the value of the total
correlation function. One has $E(\vec{n}_{k_1},\vec{n}_{k_2},
\vec{z},...,\vec{z})=E(\vec{n}_{k_1},\vec{n}_{k_2})$. Therefore within the
local realism these correlation functions have to satisfy the CHSH
inequality. Of course the generalized GHZ state (\ref{STATE}) violates it
for the whole range of $\alpha\neq 0$.

What are the reasons for the completely different behavior for $N$
even and $N$ odd? The expression on the left-hand side of
(\ref{ban}) can be understood as a "total measure of the strength
of correlations" in mutually complementary sets of local
measurements (as defined by the summation over $x$ and $y$)
\cite{essence}. Then the unity on the right-hand side of
(\ref{ban}) is the classical limit for the amount of correlations.
Specifically, pure product states cannot exceed the limit of 1, as
they can show perfect correlations in one set of local
measurements directions only. In contrast, entangled states can
show perfect correlations for more than one such set
\cite{essence}. Now, only if $N$ is even, the states (\ref{STATE})
already show perfect correlation between measurements along
$z$-directions (as the product is then always +1) reaching
therefore the classical limit. Yet, they also show additional
correlations in other, complementary, directions. In the case of
$N$ odd, however, there is no perfect correlation between
measurements along z-directions and the amount of correlations in
complementary directions are not enough to violate (\ref{ban}).

In summary, we have shown that Gisin's theorem cannot be
straightforwardly generalized to all multi-particle systems, for
the case of Bell inequalities involving correlation functions, in
which local observers can choose between two dichotomic
observables (see Statement 1 and 2). However, the question posed
in the title of this letter may find a different answer, if, e.g.,
more than two dichotomic observables per local measurement station
are allowed.

Our results may shell a new light on the connection between the
violation of Bell inequalities and the quantum information tasks
\cite{ScaraniPRL,acin}. In the problem of classification of
entangled states our results reveal a new class of pure entangled
states which do not violate any Bell inequalities for correlation
functions.

\v{C}.B. is supported by the Austrian FWF, project F1506, and by
the QIPC Program of the European Union. The work is a part of the
Austrian-Polish  program 24/00. M.\.{Z}. acknowledges KBN grant
No. 5 P03B 088 20. W.L. and M.W. are supported by the University
of Gda\'nsk grant BW 5400-5-0236-2. Anton Zeilinger and Jian-Wei
Pan are warmly thanked for discussions.

%\vspace{-0.1cm}

\end{document}